# Electron doping induced magnetic glassy state in phase separated YBaCo$_2$O$_{5.5-\delta}$

Archana Kumari*, C. Dhanasekhar and A. K. Das

*Department of Physics, Indian Institute of Technology, Kharagpur -721302, India*

*e-mail: tiwariarchana9@gmail.com

**ABSTRACT**

The structural, magnetic and transport properties of the layered RBaCo$_2$O$_{5.5\pm\delta}$ cobaltites are sensitive to the oxygen stoichiometry. In this present study, we report the presence of a low-temperature magnetic glassy state in electron-doped polycrystalline YBaCo$_2$O$_{5.5}$ cobaltite. The ac magnetization studies show the absence of conventional spin glass features, while the various dc magnetization studies demonstrate the presence of non-equilibrium magnetic glassy state at low temperature. The magnetic glassy state of this sample results from the kinetic arrest of the first order ferro (or ferri) to antiferromagnetic transition. The role of electron doping in the occurrence of magnetic glassy phenomena is discussed in terms of magnetic phase separation involving the Co$^{3+}$/Co$^{2+}$ clusters of the ferrimagnetic phase in the Co$^{3+}$/Co$^{3+}$ antiferromagnetic matrix.

## 1. INTRODUCTION

Layered cobaltites RBaCo$_2$O$_{5+\delta}$ (R = lanthanide ion or Y and δ ~ 0.5) have been intensively studied over the years due to unusual structural, magnetic and electronic phase transitions [1], [2], [3], [4], [5], [6], [7], [8], [9], [10]. Importantly, these cobaltites shows various transitions with decreasing temperature, starting with the metal-insulator (T$_{MIT}$ = 330 – 360 K) transition, followed by paramagnet-ferromagnet (T$_C$ = 280 – 310 K) and finally ferromagnet-antiferromagnet (T$_N$ = 200 – 250 K) transitions [6]. The all above transitions and structural properties of these cobaltites are strongly influenced by the various factors: R-cation size, the structural disorder introduced by chemical substitutions at the distinct R, Ba, and Co sites and by a variation of the oxygen stoichiometry [4], [6], [11], [12], [13], [14], [15], [16]. The layered structure with oxygen content around δ = 0.5 is extensively studied in this family which consists of the RO$_\delta$ − CoO$_2$ − BaO − CoO$_2$ sheets along the c axis [12], [13]. The oxygen stoichiometry (δ) in these compounds can be modified in a wide range (0 ≤ δ ≤ 0.9) depending on the R cation and the synthesis conditions, which follows the change of crystal structure from the tetragonal with *P4/mmm* symmetry for δ = 0, to an orthorhombic crystal structure with *Pmmm* symmetry for δ = 0.5 and in case of δ ≈ 0.9 various crystal structures i.e. tetragonal (*P4/mmm*), cubic (*Pm$\bar{3}$m*), orthorhombic (*Pbnm*) etc., are reported which strongly depend on the rare earth and barium site elements [2], [17], [18], [19], [20].

The variation of δ nearby 0.5 leads to the changes in the formal valence of Co ions; the hole doping induces the Co$^{4+}$ states (δ > 0.5) while electron doping (δ < 0.5) induces the Co$^{2+}$ states; whereas in both doping cases there exists a Co$^{3+}$. Taskin et al. have shown that any deviation from δ = 0.5 leads to a decrease in all observed critical phase transition temperatures [13]. Specifically, the T$_{MIT}$ was very slightly suppressed on the electron-doped side (δ < 0.5) and little more on the hole-doped side (δ > 0.5). Both electron and hole doped studies show that the small and large suppression of T$_C$ and T$_N$ respectively. Additionally, the magnetic and transport measurements showed the formation of two insulating phases for the electron-doped region (δ < 0.45), whereas the hole-doped region (δ

> 0.45) showed the coexistence of insulating and metallic phases. More importantly, Miao et al. studies on hole doped $PrBaCo_2O_{5.5+\delta}$ ($\delta = 0.24$) showed a large magnetovolume effect, which is due to the inserting of ferromagnetic clusters into the antiferromagnetic matrix [5]. The $\delta \approx 0.5$ cobaltites exhibit a $Co^{3+}$ high spin (HS; $t_{2g}^4 e_g^2$, S=2) – intermediate spin (IS; $t_{2g}^5 e_g^1$, S=1) and IS - low spin (LS, $t_{2g}^6 e_g^0$, S = 0) state transitions phases [1], [21], whereas both electron and hole doped cobaltites show melting of spin-state ordering [22].

Among the unusual features observed with variation of δ in these cobaltites, much emphasis has been placed on understanding the low temperature magnetic properties of the δ < 0.5 cobaltites. In this study, we report the presence of a magnetic glass-like state at low-temperature in $YBaCo_2O_{5.5-\delta}$ (δ=0.13) cobaltite. The ac magnetization studies show the absence of conventional spin glass (SG) feature, while the various dc magnetization studies, such as cooling and heating in unequal fields (CHUF), magnetization relaxation and temperature cycling magnetization measurements show the signatures of a non-equilibrium magnetic glassy state at low-temperature. The role of electron doping in the incidence of magnetic glassy phenomena is discussed in terms of the evolution of phase separation between the ferromagnetic and antiferromagnetic clusters, which is trivial in the $YBaCo_2O_{5.5-\delta}$ (δ ~ 0) cobaltite.

## 2. EXPERIMENTAL

Polycrystalline $YBaCo_2O_{5.5}$ sample was prepared by the conventional solid-state reaction method similar to the literature [15], [23]. The starting materials of $Y_2O_3$, $BaCO_3$, $Co_3O_4$, were heated at ~ 700 °C for 6 hrs to remove the moisture, etc. Stoichiometric amounts of the reactants were intimately ground and heated at 950 °C in air for 12 h for decarbonation. The mixture was then reacted in air at 1000 °C for 24 h and slowly cooled to room temperature. The mixture was reground, pressed into pellets and finally fired at 1000 °C for 24 h. This process was repeated several times until the single phase is formed, and after each annealing, the sample was slowly cooled to room temperature. The sample $YBaCo_2O_{5.5-\delta}$ was obtained by heating the $YBaCo_2O_{5.5}$ sample in pure argon gas up to 560 K and then cooling it to room temperature at the rate of 1 K/min [23]. The phase purity of the sample was checked by room temperature X-ray diffraction (XRD) using Cu Kα radiation in the 2θ range of 10º to 80º and the structural parameters were obtained using Rietveld refinement through FULLPROF program. The oxygen content (δ = 0.13) was determined by iodometric titration, which allowed the sample to be formulated as $YBaCo_2O_{5.37}$ (YBCO), and the corresponding average cobalt valence is +2.87, suggesting that $Co^{2+}$ is present together with $Co^{3+}$. Magnetization versus temperature, i.e. M (T), isothermal magnetization, i.e. M (H) and time dependent dc magnetization, i.e. M (t) measurements were carried out using Physical Property Measurement System (PPMS), Cryogenic Limited, UK. The temperature dependent ac magnetization measurements, i.e. χ (T) and temperature cycling of magnetization measurements were performed using a superconducting quantum interference device (SQUID) magnetometer, Quantum Design, San Diego, USA.

## 3. RESULTS AND DISCUSSION

### 3.1 Structural Characterization

The experimental XRD profile at room temperature together with the profile calculated from the Rietveld refinement of the polycrystalline powder sample $YBaCo_2O_{5.37}$ (YBCO) in the range of $10° \leq 2\theta \leq 80°$ is shown in Fig. 1 (a). The Rietveld analysis shows the tetragonal symmetry with *P4/mmm* space group ($a_p \times a_p \times 2a_p$; where $a_p$ is the lattice parameter of a simple perovskite) and the obtained lattice parameters are a = 3.872 (2) Å, b = 3.872

(2) Å, c = 7.544(7) Å and unit cell volume = 116.5(4) Å$^3$ [23]. The valence states of the Co ion of YBaCo$_2$O$_{5.37}$ sample was characterized by XPS measurements and the corresponding Co 2p spectra are shown in Fig. 1(b). The binding energy and oxidation states were estimated in comparison with adventitious carbon of binding energy 284.55 eV. The core-level spectra of Co 2p are classified into two regions: Co 2p$_{3/2}$ and Co 2p$_{1/2}$ due to spin-orbit splitting (15.33 eV). The shake-up satellite peaks for the Co 2p$_{3/2}$ and Co 2p$_{1/2}$ peaks due to ligand-to-metal charge transfer processes are detected at higher binding energies compared to the main peaks. The main and satellite peaks of Co 2p$_{3/2}$ are observed at 779.30 and 789.07 eV, respectively, while the main and satellite peaks of Co 2p$_{1/2}$ are observed at 794.65 and 805.0 eV, respectively. By using a Lorentzian-Gaussian fitting method, the Co 2p spectra was further deconvoluted into two separate peaks: one set is located at 779.08 and 794.35 eV for Co$^{2+}$, while another set is located at 780.21 and 795.32 eV for Co$^{3+}$.

**3.2 Magnetization studies**

The dc magnetization vs temperature curves of YBCO sample recorded under magnetic field of 0.1 T using zero-field cooling (ZFC), field-cooled-cooling (FCC) and field-cooled-warming (FCW) cycles are shown in the Fig. 2 (a) in the temperature range 5-390 K. As the temperature decreases, a slope change is observed in the M (T) curve at T ~ 360 K (see right inset of Fig. 2(a)), which might be due to the spin state transition of Co$^{3+}$ (HS-IS/LS) and is observed for other cobaltites [6]. Upon further lowering of the temperature, the magnetization increases sharply below 300 K and then drops below 190 K. The magnetic transition is very broad and this behavior is similar to those seen in the sample with δ < 0.10 [23]. T$_C$ and T$_N$ are obtained from the derivative of FCC magnetization curve and are found to be 293 K and 160 K respectively. In case of δ ≈ 0.50 sample, the transitions are reported at 297 K (T$_C$) and 270 K (T$_N$) [23], [24]. However, in the present δ ≈ 0.37 case both the transitions are shifted to low temperature, which indicates that the electron doping plays an important role and this behavior corroborates with the literature [23].

Further, the YBCO sample shows a thermal hysteresis between the FCC and FCW curves in the temperature range of 270 - 100 K, and for T < 100 K, both the FCC and FCW curves overlap which is shown in the Fig. 2 (a). This behavior is in contrast to the YBaCo$_2$O$_{5.5}$, which does not show any significant thermal hysteresis between the FCC and FCW magnetization curves from 350 - 5K [25], [26]. The thermal hysteresis is a usual common characteristic feature of the first order phase transition (FOPT) and is observed in various other cobaltites of this family and also other materials which lead to supercooling, and phase coexistence that can be eventually termed as a magnetic-glass phase [27], [28], [29], [30]. In addition to thermal hysteresis, the YBCO shows large thermomagnetic irreversibility (TMI) between ZFC and FCC/FCW curves, which increases at low temperature. The TMI at T = 10 K was calculated from the difference between the ZFC and FCC magnetization values i.e., ΔM = M$_{FCC (T = 10 K)}$ − M$_{ZFC (T = 10 K)}$ and ΔM is found to be increased with the applied magnetic field (inset of Fig. 2 (a)). This is different from conventional magnetic systems, where ΔM actually decreases with an increase in the magnetic field. In this family, number of materials shows similar TMI at low temperature and is attributed to the direct consequence of phase coexistence [26], [30]. The magnetization value of the present YBCO samples is slightly greater than that of the YBaCo$_2$O$_{5.50}$ (which is purely antiferromagnetic) at low temperature [25], [26]. The M (H) measured at 5 K shows a straight line for the case of YBaCo$_2$O$_{5.50}$ [25], whereas the present sample shows significant hysteresis which indicates the coexistence of both AFM and FM phases below T$_C$ (Fig. 2 (b)).

The magnetic ground state of the YBCO sample is further investigated using ac magnetization studies under a small ac field of 1 Oe and at different frequencies. The results obtained for the in-phase ac magnetization component, ($\chi'_{ac}$) and the out-of-phase ac magnetization component, ($\chi''_{ac}$) vs. temperature are presented in Fig. 3. The $\chi'_{ac}$ shows an abrupt sharp increase in the vicinity of $T_C$ = 240 K and slight decrease with temperature up to $T_N$. For T < $T_N$ the $\chi'_{ac}$ falls very rapidly and below 100 K it is found to be independent of the measured frequency. Although the $\chi'_{ac}$ does not show any significant frequency shift at $T_C$ and $T_N$ but at every temperature in between 100 K – 240K the $\chi'_{ac}$ magnitude slightly decreases with increasing frequency. This effect is observed near the FM–AFM transition and also where a significant thermal hysteresis between the FCC and FCW curves are present. On the other hand the $\chi''_{ac}(T)$ shows a frequency independent peak at 240 K, which is the onset of the AFM transition. Similar behavior also observed in $YBaCo_2O_{5.50}$ [31] and other cobaltites in this family [32].

The $\chi'_{ac}$ behavior is consistent with the behavior of dc magnetization, and these ac magnetization studies exclude the SG like feature in the YBCO sample. On the other hand, the dc magnetization measurements show signatures of magnetic glassy (MG) like state at low temperature. The MG state in YBCO is initially studied using a cooling and heating in unequal field (CHUF) experimental protocol [30], [33]. In this CHUF measurements, the sample is cooled from high temperature (T > $T_C$) to lowest temperatures under a certain applied magnetic field ($H_{cool}$). At the lowest temperature, $H_{cool}$ is isothermally converted to a different measuring field ($H_{measure}$) which may be larger or smaller than $H_{cool}$, and the magnetization is recorded while warming the sample in the presence of $H_{measure}$. The obtained CHUF results of the YBCO samples are shown in the Fig. 4. For T > $T_C$ the magnetization behavior of YBCO under the CHUF protocol remains the same irrespective of the cooling field. For T < $T_C$, the magnetization curves show distinct behavior and can be categorized into two groups based on their low-temperature behavior. For $H_{cool}$ ≤ $H_{measure}$ [curves 1–3 in Fig. 4], the magnetization is almost constant at low temperature and shows a sharp rise above 120 K [marked by the black arrow]. While for $H_{cool}$ > $H_{measure}$ [curves 4 and 5 in Fig. 4] the magnetization decreases gradually up to 120 K and starts increasing above 120 K [marked by the two pink arrows]. This behavior can be realized as follows: irrespective of the $H_{measure}$ the magnetization increases above 120 K, which is due to the transition from antiferromagnetic to ferromagnetic state. For $H_{cool}$ > $H_{measure}$, the FM component is frozen (or kinetically arrested) on the AFM background at low temperatures and acts as a glass which is devitrified in the subsequent heating cycle and approaches the equilibrium AFM phase, resulting in a rapid decrease in magnetization. On the other side for $H_{cool}$ ≤ $H_{measure}$ the fraction of frozen FM component is small and unable to overcome the energy barrier to approach the equilibrium AFM phase.

Further, to understand the low temperature magnetic glassy feature in YBCO sample, temperature-cycling in ZFC mode is performed [27], [33], [34] and the obtained results are shown in Fig. 5. In this experiment, the sample is cooled from 350 K to 10 K (≈$T_0$) without any magnetic field. At $T_0$, a magnetic field of 0.1 T was applied and magnetization was recorded while the sample was heated ($T_0$ - $T_W$) and cooled ($T_W$ - $T_0$) at a certain temperature interval. Starting with $T_W$ = 20 K this temperature cycling is continued with 10 K temperature interval from 20 K-120 K. Magnetization at $T_0$ clearly increases after the first temperature cycle and this rise in magnetization in the subsequent cycles becomes larger as $T_W$ becomes higher. The magnetization measured during cooling and heating follows the same path up to 90 K. However, at each $T_W$ the magnetization value increases continuously from that of the previous one. This observation clearly indicates that even at a fixed magnetic field, at low temperatures, the thermal cycling can convert part of the supercooled, metastable FM phase into an equilibrium AFM phase. When $T_W \geq 90$ K, the initial increase in magnetization becomes almost discontinuous, and the thermal hysteresis between cooling and heating cycles is detected above 90 K, as shown in Fig. 5. This result shows that below 90 K the thermal cycling gradually increases the nucleation and growth of the FM state from the mixed FM and AFM states.

To explore the low temperature non-equilibrium magnetic state, we measured the magnetic relaxation under FC experimental condition. In FC mode, the sample was cooled from 380 K to 20 K in the presence of an applied magnetic field of 2 T, then the sample was held at 20 K for a waiting time of 300 s, after which the magnetic field was removed and the magnetization was recorded as a function of time. The normalized time dependence of the magnetization, i.e. $M(t)/M_0$ (where t and $M_0$ are the relaxation time and magnetization at $t = 0$ respectively) is shown in the Fig. 6. The time evolution of magnetization increases more rapidly when the measurement temperature is increased from 20 K to 100 K while it decreases at temperatures above 100 K. This FC magnetization relaxation can be explained as follows: at low temperatures, there is a free energy barrier that separates the supercooled FM phase from the equilibrium AFM phase and causes the lower magnetization relaxation. As the temperature increases, the supercooled FM phase starts to relax due to the thermal activation. Further, increasing temperature the magnetic relaxation is affected by the competition between the thermal energy of the supercooled FM phase and the energy barrier between the FM and AFM phases, which might cause a slower relaxation at T = 150 K and T = 175 K even though the thermal energy is higher.

The normalized time dependence of the magnetization is analysed using Kohlrausch-Williams-Watt (KWW) stretched exponential function $\varphi(t) \propto \exp\left[-\left(\frac{t}{\tau}\right)^\beta\right]$, where $\tau$ is the characteristic relaxation time and $\beta$ is a shape parameter [27], [28]. The inset of Fig. 6 shows a KWW stretched exponential function fitting at $T = 20$ K and the obtained $\tau = 1.1 \times 10^7$ s, and $\beta = 0.45$ respectively. A similar fitting at other temperatures show the value of $\beta$ varies between 0.45 and 0.7. Further, the relaxation time $\tau$ values are found to be in the range of $10^9$ s to $10^7$ s. For glassy systems, $\beta$ remains in the range of $0 < \beta < 1$ and various studies shows that the relaxation time $\tau$ values are higher in the magnetic glassy systems as compared to the conventional spin glass (SG) [27], [28], [35], [36]. The obtained $\beta$ values signify that multiple anisotropy barriers are distributed in the present YBCO and the relaxation evolving through a number of intermediate metastable states. The similar kind of magnetization relaxation has been observed in the $Gd_5Ge_4$ [27], Ce $(Fe_{0.96}Ru_{0.04})_2$ [28] and Heusler compound $Mn_2PtGa$ [37], which shows a larger relaxation time $\tau$ values ($10^5$-$10^9$) and $\beta$ is found to be in the range of 0.1 - 0.8.

The observed magnetization dynamics in oxygen deficient (or electron doped) $YBCo_2O_{5.5}$ is relatively similar to that of 5% calcium-doped (at Ba or Y sites) $YBaCo_2O_{5.5}$ cobaltite. The latter compound also undergoes PM-FM-AFM transitions and magnetic glassy state at low temperature. More interestingly, it has been shown that the magnetic glassy properties of $YBCo_2O_{5.5}$ is independent of the site of Ca substitution, either Y or Ba site. [30]. This is quite surprising and to explain the identical magnetic properties of 5% calcium (at Ba or Y sites) doped $YBaCo_2O_{5.5}$ cobaltite it is assumed that $Co^{3+}$ disproportionation into $Co^{4+}$ and $Co^{2+}$, which causes the local phase separation between the ferro- or ferrimagnetic ($Co^{4+}$) clusters and antiferromagnetic ($Co^{2+}$) clusters. Thus, this local phase separation leads to a magnetic glassy state. Note, in the Ca doped $YBaCo_2O_{5.5}$ cobaltite samples the oxygen stichometry is close to the $O_{5.5}$, whereas in the present case, we have shown the formation of magnetic glassy state in the electron doped YBaCo$_2$O$_{5.5}$ cobaltite. Further, the above XPS studies show the electron doped YBaCo$_2$O$_{5.5}$ cobaltite containing both divalent (Co$^{2+}$) and trivalent (Co$^{3+}$) cobalt ions. The presence of Co$^{3+}$/Co$^{2+}$ clusters of the ferri (or ferro) magnetic phase in the Co$^{3+}$/Co$^{3+}$ AFM matrix [38] can give the magnetic glassy state. The magnetic glass state in electron and Ca-doped YBaCo$_2$O$_{5.5}$ cobaltite differs from other magnetic glass systems such as the alloy Gd$_5$Ge$_4$ [27] and Ce(Fe$_{0.96}$Ru$_{0.4}$)$_2$ [28]. In the case of Ce(Fe$_{0.96}$Ru$_{0.4}$)$_2$, the magnetic glassy state is developed only above a certain critical magnetic field, while in Gd$_5$Ge$_4$ the magnetic glassy state is observed at low magnetic fields.

## 4. CONCLUSIONS

In conclusion, we have studied the phase coexistence and metastable magnetic glassy behavior in the electron doped YBaCo$_2$O$_{5.5}$ cobaltite. This glassy state arises from the kinetic arrest of a first-order ferro (or ferri) magnetic to the antiferromagnetic phase transition. Cooling the sample from a temperature above the ferromagnetic transition induces a supercooled ferromagnetic phase in an antiferromagnetic background, which prevents the system from reaching its magnetic ground state. In layered cobaltites, a number of compounds showing first-order ferro- (or ferri) magnetic - antiferromagnetic transition and phase separation behavior, can be potential sources for magnetic glassy systems. More importantly, oxygen stoichiometry decides the various properties of these materials. The present study highlights the formation of magnetic glassy state in the oxygen deficient YBaCo$_2$O$_{5.5}$ cobaltite.

**ACKNOWLEDGMENTS**

The authors acknowledge the use of PPMS and XPS facility under DST-FIST project in the department of physics and central research facility, IIT Kharagpur.

**Figure captions**

**FIG.1. (a)** *Room temperature XRD pattern of the polycrystalline $YBaCo_2O_{5.37}$ sample with Rietveld refinement analysis. Different colour indicates the experimental data (red circles), the Rietveld data (blue lines), the difference between experimental and Rietveld data (black lines) and the Bragg's peaks position (green vertical lines)* **(b)** *Deconvoluted high-resolution Co 2p spectra.*

**FIG.2.** *(a) Magnetization vs temperature curves for $YBaCo_2O_{5.37}$ following the ZFC, FCC, and FCW measurement protocols under a magnetic field of H = 0.1 T. Inset shows the variation of $\Delta M$ (= $M_{FCC\,(T\,=\,10\,K)}$ − $M_{ZFC\,(T\,=\,10\,K)}$) as a function of magnetic field (left) and spin state transition (right) (b) M (H) curve of $YBaCo_2O_{5.37}$ at 5 K.*

**FIG.3.** *In-phase ac-magnetization curve ($\chi'_{ac}$ (T)) for $YBaCo_2O_{5.37}$ measured at different frequencies under ac field of 0.1 Oe as a function of temperature. Inset shows out of phase ac-magnetization curve ($\chi''_{ac}(T)$).*

**FIG.4**. *Magnetization vs temperature for $YBaCo_2O_{5.37}$ measured using the CHUF experimental protocol. The sample is cooled in different magnetic fields ($H_{Cool}$) and the magnetization is measured while warming the sample in the measuring field ($H_{Measure}$) of 1 T.*

**FIG.5**. *Magnetization vs temperature plots recorded in 0.1T showing the effect of zero field cooled temperature-cycling starting from $T_0$= 10 K. The yellow box highlights the discontinuity of the magnetization and progress of thermal hysteresis at high temperature.*

**FIG.6**. *Time dependence of the normalized magnetization for $YBaCo_2O_{5.37}$ measured at different temperatures under FC protocol. Inset shows KWW stretched exponential function $\left( \varphi(t) \propto exp\left[-\left(\frac{t}{\tau}\right)^\beta\right] \right)$ fitting at 20 K.*

**Figures**

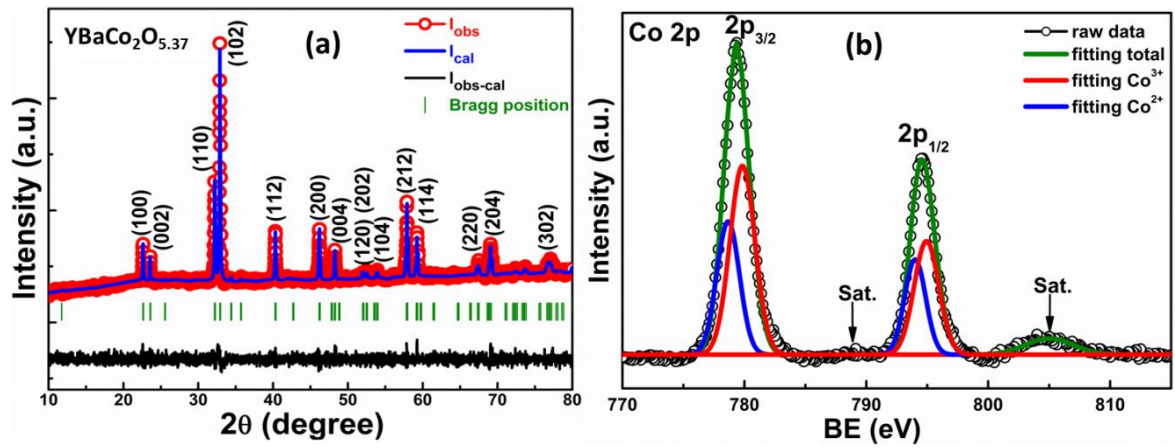

**FIG. 1**

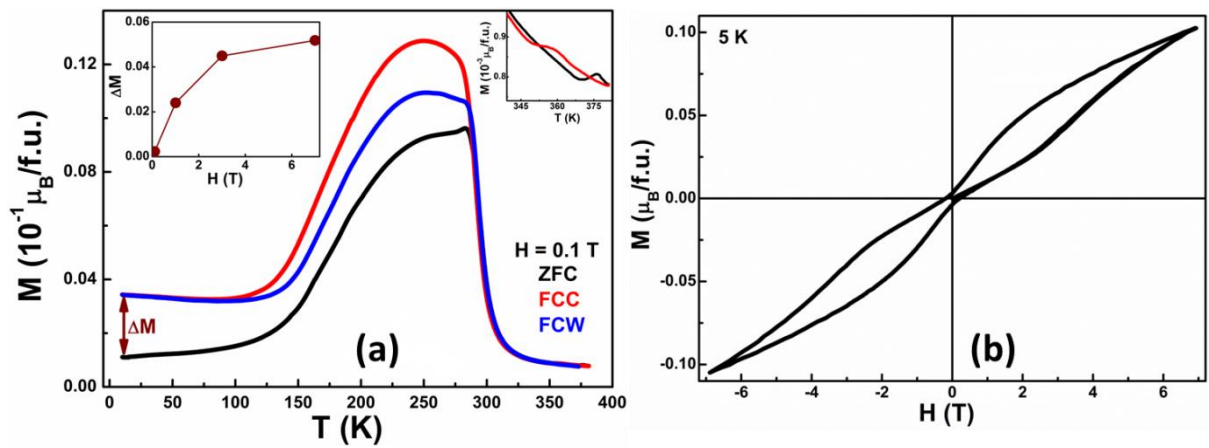

**FIG. 2**

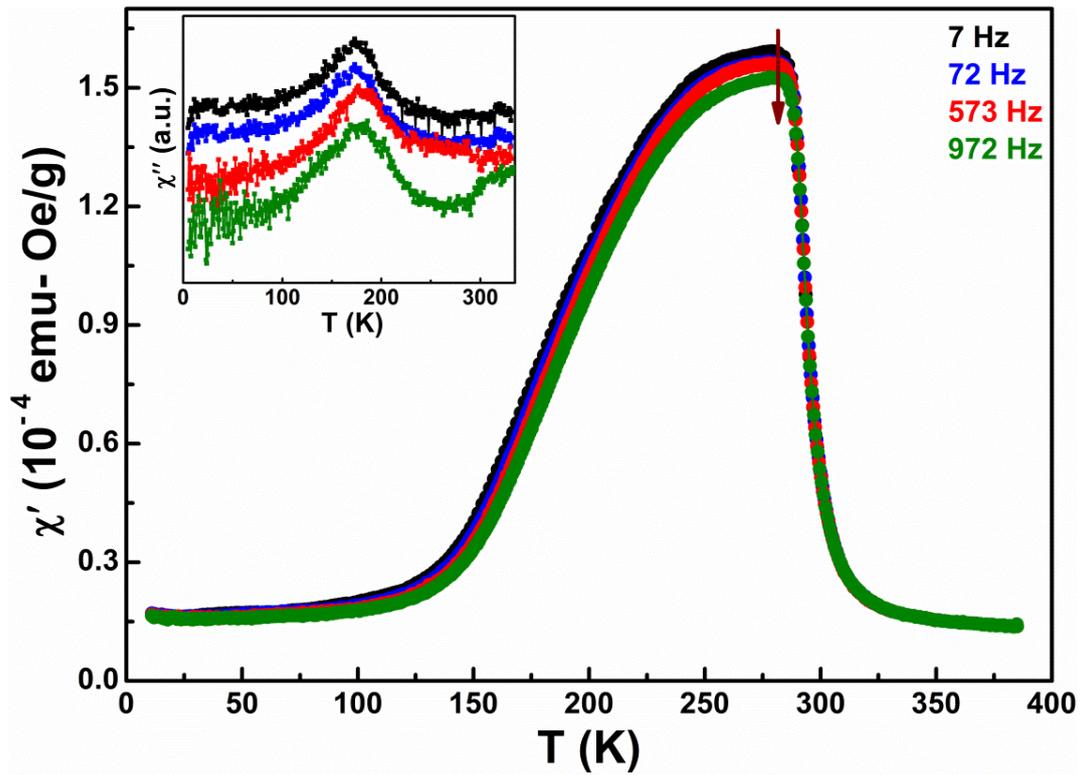

**FIG. 3**

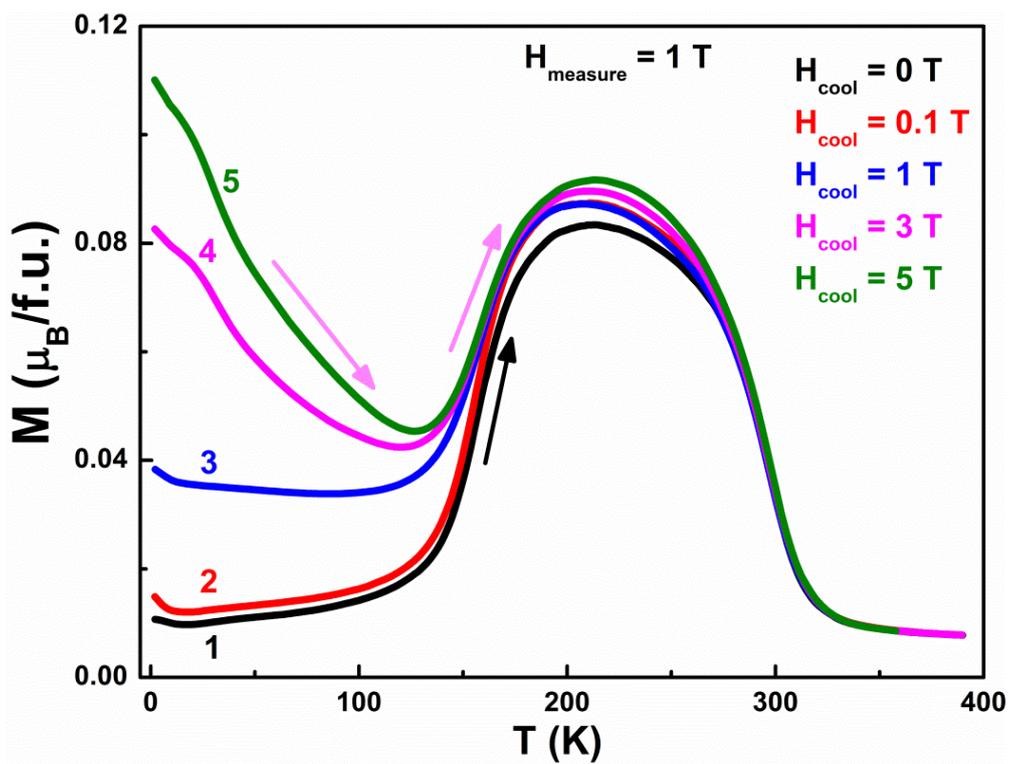

**FIG. 4**

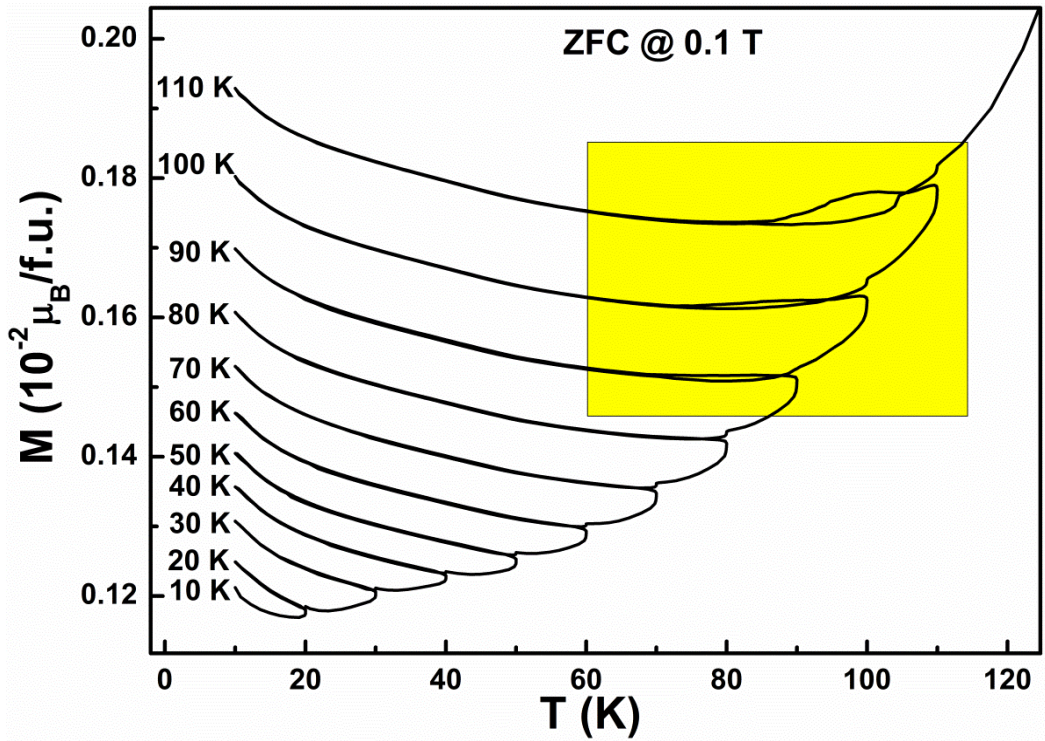

**FIG. 5**

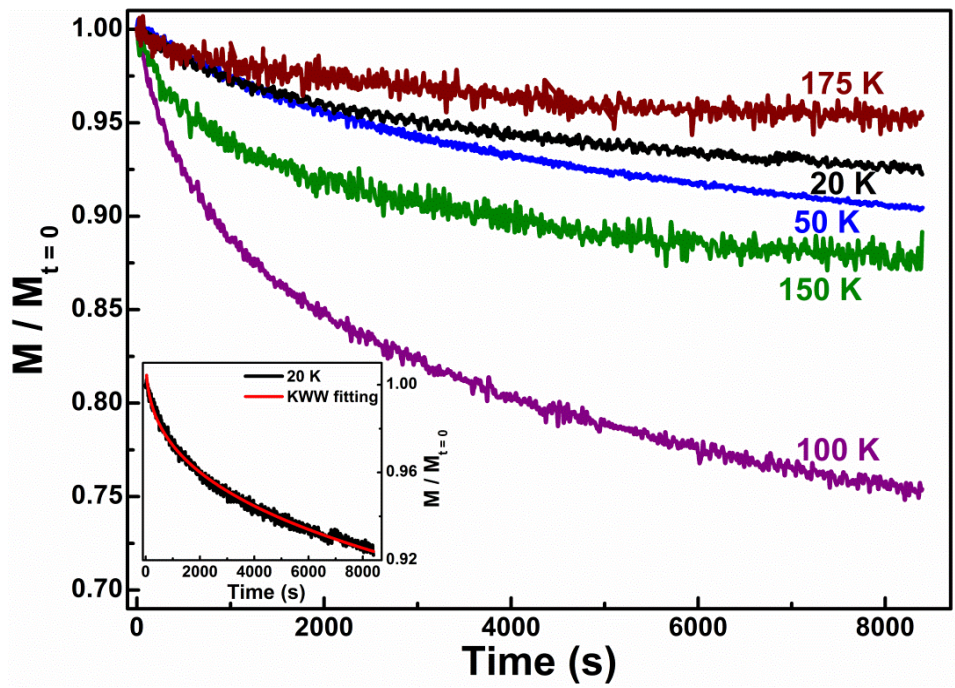

**FIG. 6**